\documentclass[prb,groupedaddress, showkeys,showpacs]{revtex4}

\usepackage{graphicx}
\usepackage{amsmath}
\usepackage{mathrsfs}

\begin{document}


\title{Sorting of chiral active particles driven by rotary obstacles}
\author{Qun Chen }
\author{Bao-quan Ai}  \email[Email:]{aibq@scnu.edu.cn}
\affiliation{ Guangdong Provincial Key Laboratory of Quantum Engineering and Quantum Materials,
School of Physics and Telecommunication Engineering,
South China Normal University, 510006 Guangzhou, China.}


\date{\today}
\begin{abstract}
  \indent Sorting of microswimmers based on their mobility properties is of utmost importance for various branches of science and engineering. In this paper, we proposed a novel sorting method, where the mixed chiral particles can be separated by applying two opposite rotary obstacles. It is found that when the angular speed of the obstacles, the angular speed of active particles and the self-propulsion speed satisfy a certain relation, the  mixed particles can be completely separated and the capture efficiency takes its maximal value. Our results may have application in capture or sorting of chiral active particles, or even measuring the chirality of active particles.
  \end{abstract}

\pacs{05. 40. -a, 82. 70. Dd, 05. 60. Cd}



\maketitle
\section {Introduction}
\indent In recent years a significant amount of research has been focused on active particle systems, where the internal energy or energy from the local
environment can be converted into the directed motion\cite{rmp,Reimann,Ebbens,rmp1,rpp}. Compared with passive Brownian particles, active Brownian particles are able to consume internal energy and actively facilitate directed motion in the out-of-equilibrium environment\cite{Taylor,Schweitzer,Erdmann,Toner,Vicsek1}. The motion of active particles not only depends on a random diffusion, but also is determined by a self-propelling force and a torque (if considering the chirality of active particles)\cite{Berg,Purcell,Howse,Teeffelen,Leonardo,Aguilar,Yang,Wensink,H. U.,Friedrich}. The dynamics of active particles could exhibit peculiar behaviors\cite{Golestanian,Vicsek,Ramaswamy,Peruani,Giomi,Saintillan,Shen,Burada,Tailleur,Cates,Angelani,Fily,Kaiser,Ghosh,angelani,Buttinoni,Nguyen},    such as anomalous density fluctuations\cite{Ramaswamy}, clustering\cite{Peruani}, unusual rheological behavior\cite{Giomi,Saintillan}, and activity-dependent phase boundary changes\cite{Shen}.

\indent Active particles can be captured and separated based on their mobility properties\cite{Wan,Tailleur1,WYang,McCandlish,Maggi,Costanzo,Y,X,Berdakin,Reichhardt,Mijalkov,Volpe,Ai1}. Yang and his coworkers\cite{WYang} found that particles in a binary mixture can be effectively separated by using self-driven particles. McCandlish and his coworkers\cite{McCandlish} found that the difference in collision frequencies between passive and active particles can provide a driving force for separating passive and active particles. Maggi et al.\cite{Maggi} demonstrated that the centrifugation induces significant effect on the separation of active particles with different motilities. Costanzo\cite{Costanzo} studied the separation of run-and-tumble particles in terms of their motility in a channel. Fily et al.\cite{Y} showed that the boundary shape of containers affects the dynamics of active fluid and particles can be trapped at the boundary. Yang et al.\cite{X} found that a mixture of bidisperse disks could be segregated in a confined box. In the environment with asymmetric obstacles, Berdakin and his coworkers\cite{Berdakin} proposed different swimming strategies for active particles separation. Reichhardt\cite{Reichhardt} separated particles with different chirality by using asymmetric patterned arrays. Based on the chirality of active particles, active particles with opposite chirality can be separated in the chiral flower \cite{Mijalkov}. In the channel with rigid half-circle obstacles, the mixed chiral particles can be separated by applying a contain force or a shear flow\cite{Ai1}.

\indent Most studies on particle separation have referred to the static obstacle. However, in some real and artificial systems, the rotary obstacles may play an important role in transport. Therefore, it would be interesting to sorting particles in a system with rotary obstacles. In this paper, we proposed a new sorting method, where the opposite chiral particles can be captured and separated by applying two opposite rotary obstacles. We found that when the angular speed of the obstacles, the angular speed of active particles and the self-propulsion speed satisfy a certain relation, the mixed particles can be completely separated and the capture efficiency takes its maximal value.

\section{Model and methods}
\begin{figure}[htbp]
\begin{center}\includegraphics[width=10cm]{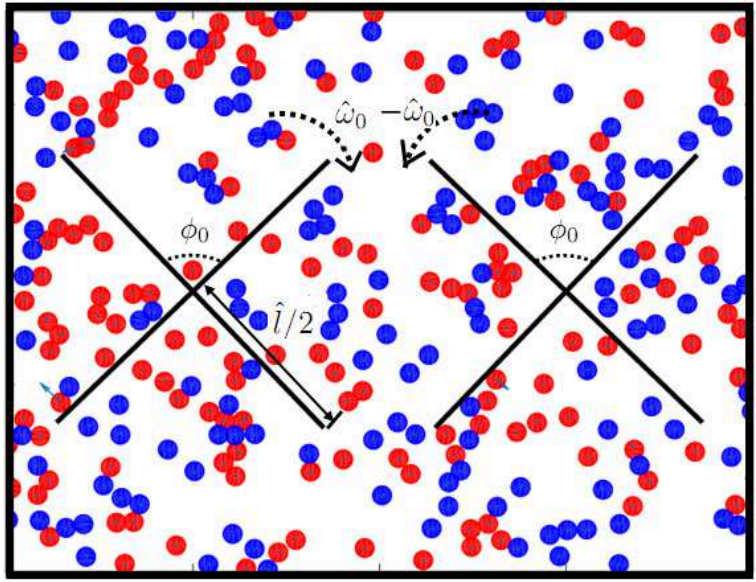}
\caption{(Color online) Scheme of sorting device: particles with opposite chirality moving in the box, where there exist two rotary obstacles with different rotation direction($\phi_0=\frac{\pi}{2}$). The hard wall boundaries are imposed in both $x$ direction and $y$ direction. The red and blue balls denote CCW and CW particles, respectively. }\label{1}
\end{center}
\end{figure}

\indent  We consider the mixtures of chiral active particles moving in a confined box with the height $\hat{H}$ and the length $\hat{L}$. In the box, there exist two rotary obstacles which consist of two intersecting linear boundaries with the length $\hat{l}$ (as shown in Fig. 1). Two obstacles rotate with angular speed $\hat{\omega}_{0}$ and $-\hat{\omega}_{0}$, respectively. Obstacles rotate clockwise (counterclockwise) for negative (positive) $\hat{\omega}_{0}$. The dynamics of particle $i$ with radius $a$ is described by the position $\hat{\mathbf{r}}_i\equiv (\hat{x}_i,\hat{y}_i)$ of its center and the orientation $\theta_i$ of a polar axis ${{\mathbf{n}}}_{i}\equiv (\cos\theta_i,\sin\theta_i)$. The particle $i$ obeys the following overdamped Langevin equations,
\begin{equation}\label{e1}
  \frac{d\hat{\mathbf{r}}_i}{d\hat{t}}={\mathbf{n}}_i \hat{v}_0+\mu \sum_{j\neq i}\hat{\mathbf{F}}_{ij}+\mu \sum_{j\neq i}\hat{\mathbf{G}}_{ij},
\end{equation}
\begin{equation}\label{e2}
  \frac{d\theta_{i}}{d\hat{t}}=\hat{\Omega}_{i}+\sqrt{2\hat{D}_{\theta}}\hat{\xi}_{i}(\hat{t}),
\end{equation}
with $\mu$ the mobility and $\hat{v}_0$ the self-propulsion speed. $\hat{D}_{\theta}$ is the rotational diffusion coefficient. $\hat{\Omega}$ is the angular velocity of chiral active particles and the sign of $\hat{\Omega}$ determines the chirality of the particle. The trajectory bends counterclockwise (clockwise) for positive (negative) $\hat{\Omega}$. For convenience, we call them as counterclockwise (CCW)particles and clockwise(CW)particles, respectively. $\hat{\xi}_{i}(t)$ denotes the Gaussian white noise of zero mean and satisfies $\langle \hat{\xi}_{i}(t)\hat{\xi}_{j}(s)\rangle = \delta_{ij}\delta(t-s)$. Note that in many realizations ( e.g. bacterial suspensions and active colloids), the rotational noise is athermal and the effective diffusion from $\hat{D}_{\theta}$ is much larger than the thermal diffusion, so we neglect thermal noise in this paper.

\indent We consider particle-particle interaction $\hat{\mathbf{F}}_{ij}$  and particle-obstacle interaction $\hat{\mathbf{G}}_{ij}$ as spring forces with the stiffness constant $\hat{k}_{pp}$ and $\hat{k}_{po}$, respectively. $\hat{\mathbf{F}}_{ij}=\hat{k}_{pp}(2a-r_{ij})\mathbf{n}_i$, if $r_{ij}<2a$ ($\hat{\mathbf{F}}_{ij}=0$, otherwise). $\hat{\mathbf{G}}_{ij}=\hat{k}_{po}(a-r_{ij})\mathbf{n}_i$, if $r_{ij}<a$ ($\hat{\mathbf{G}}_{ij}=0$, otherwise). Here, $r_{ij}$ is the distance between the  particle $i$ and the object (particle or obstacle) $j$.  We define the ratio between the area occupied by particles and the total available area as the packing fraction $\phi=\frac{N\pi a^2}{\hat{L}\hat{H}}$, where $N$ is the total number of particles.

\indent Eqs. (1) and (2) can be rewritten in the dimensionless form  by introducing characteristic length
scale and time scale  $\mathbf{r}_i=\frac{\hat{\mathbf{r}}_i}{a}$, $t=\mu k_{pp} \hat{t}$,
\begin{equation}\label{eq3}
  \frac{d{\mathbf{r}_{i}}}{d{t}}=\mathbf{n}_i{v}_0+\sum_{j\neq i}\mathbf{F}_{ij}+\sum_{j\neq i}\mathbf{G}_{ij},
\end{equation}
\begin{equation}\label{eq4}
  \frac{d\theta_i}{d{t}}={\Omega_i}+\sqrt{2{D}_{\theta}}{\xi_i}({t}),
\end{equation}
and the other parameter are $v_0=\frac{\hat{v}_0}{\mu \hat{k}_{pp}a}$, $k_{po}=\frac{\hat{k}_{po}}{\hat{k}_{pp}}$, $\Omega=\frac{\hat{\Omega}}{\mu \hat{k}_{pp}}$, $\omega_0=\frac{\hat{\omega}_0}{\mu \hat{k}_{pp}}$, $D_{\theta}=\frac{\hat{D_{\theta}}}{\mu \hat{k}_{pp}}$, $L=\frac{\hat{L}}{a}$, $H=\frac{\hat{H}}{a}$ and $l=\frac{\hat{l}}{a}$.

\indent To quantify particle separation, we calculate the Gini coefficient,
\begin{equation}\label{g}
g=\frac{1}{N}(|N_{cw}^{L}-N_{ccw}^{L}|+|N_{cw}^{R}-N_{ccw}^{R}|),
\end{equation}
where $N_{cw}^{L}$ ($N_{ccw}^{L}$) is the number of CW (CCW) particles in the left area $x<L/2$ and $N_{cw}^{R}$ ($N_{ccw}^{R}$)is the number of CW (CCW) particles in the right area $x>L/2$. The Gini coefficient $g$ approaches 0 when the distribution of particles is homogeneous and 1 when two types of particles are completely separated.

\indent In order to describe the capture of active particle, we define the capture efficiency as $\eta=N_c/N_{cw}$, where $N_c$ is the number of CW particles in the area swept by the left rotary obstacles and $N_{cw}$ is the total number of CW particles.

 \section{Results and discussion}
\indent In our simulations, unless otherwise noted, we take $L=80.0$, $H=60.0$, $l=40.0$, $\phi=0.1$, $D_{\theta}=0.0005$, and $k_{po}=100.0$.  Note that the behavior of interest (separation of active particles) in this study is not very sensitive to local density, the influence of boundary conditions on particle separation is very small, therefore we only use the closed boundary conditions in the simulations.

\begin{figure}[htbp]
\begin{center}\includegraphics[width=8cm]{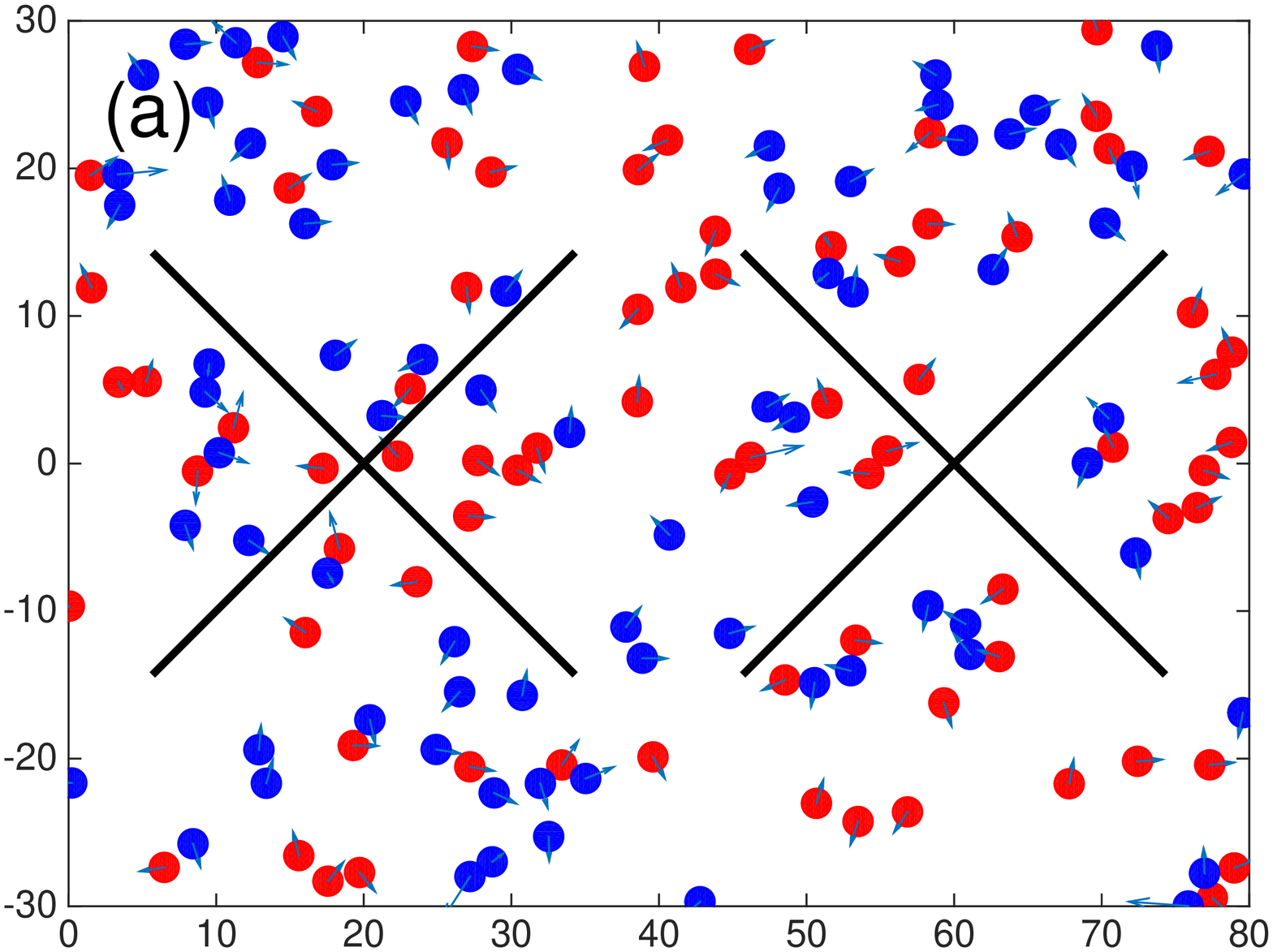}
\includegraphics[width=8cm]{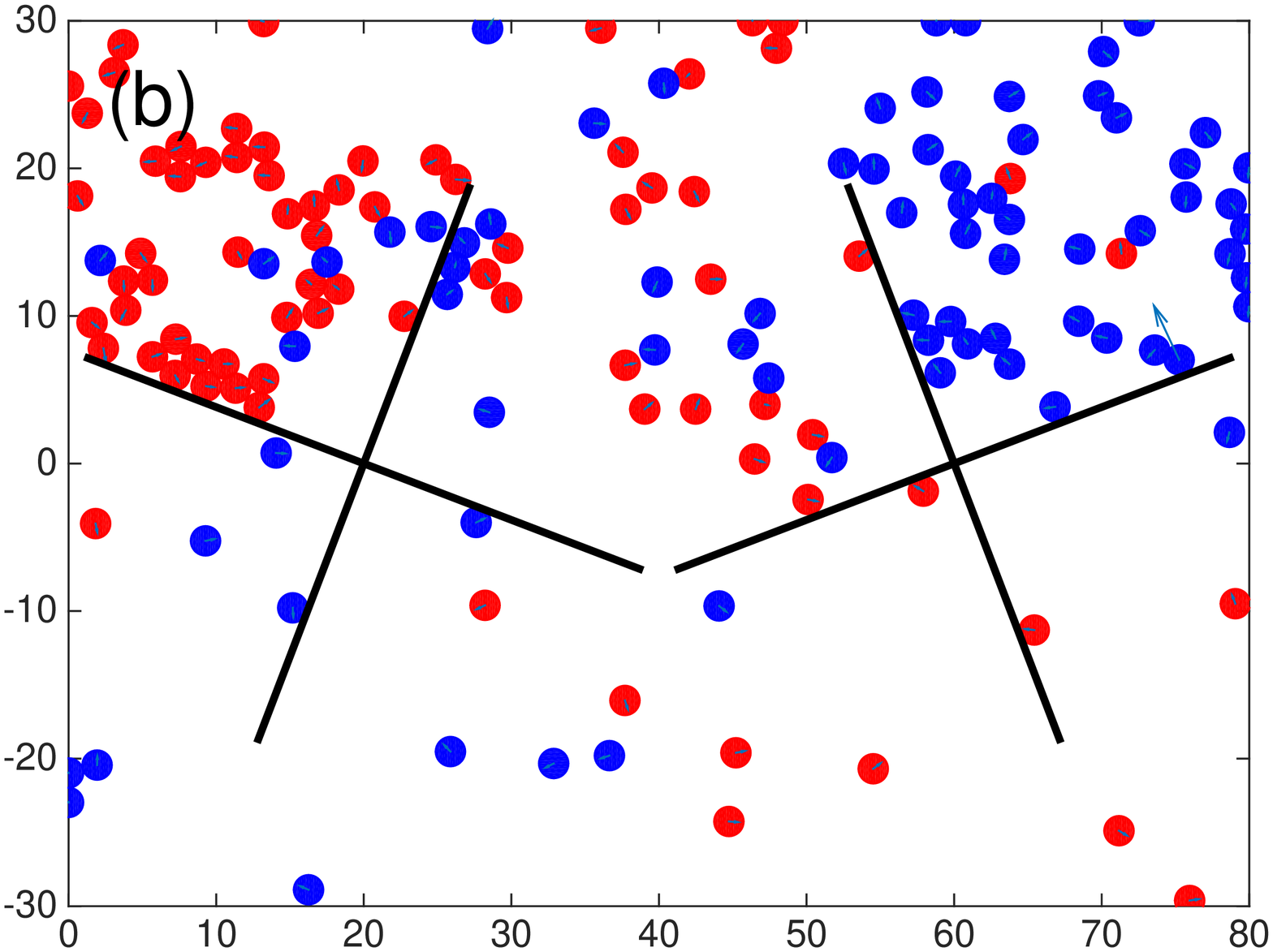}
\includegraphics[width=8cm]{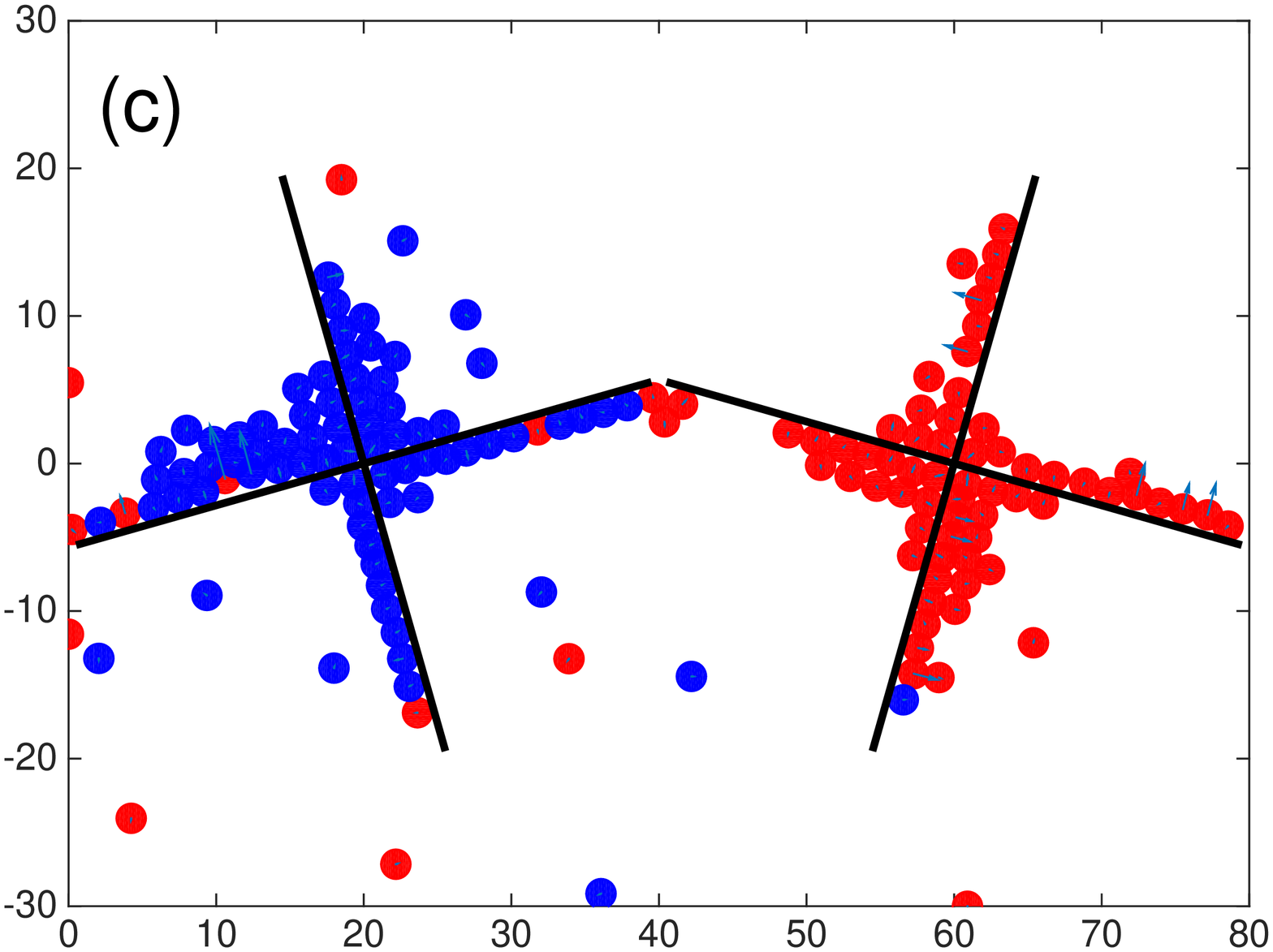}
\includegraphics[width=8cm]{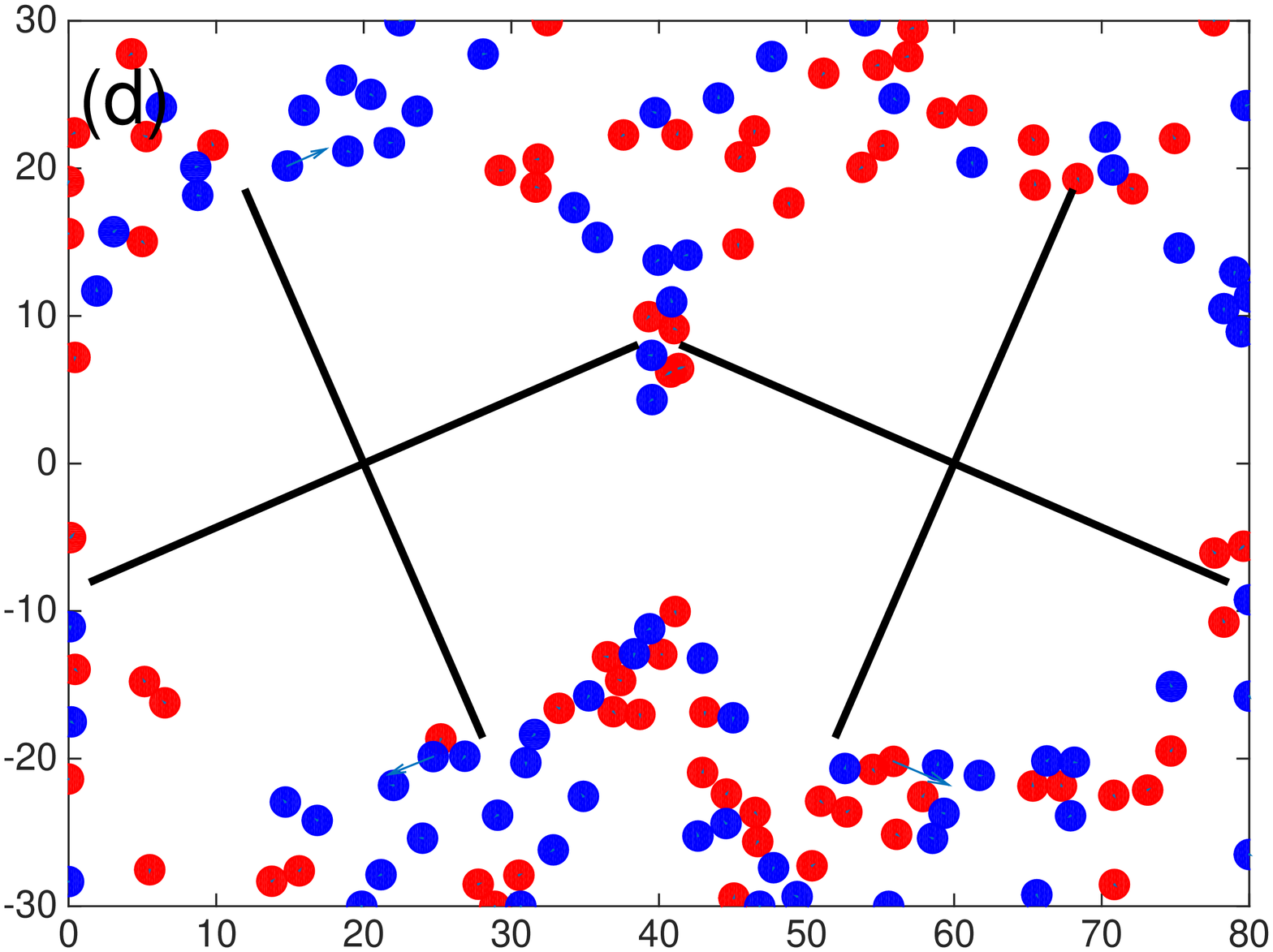}
\caption{(Color online) Snapshot of the mixed chiral particles for different values of $\omega_{0}$ at $v_{0}=0.2$ and $\Omega=0.02$. (a)at $t=0$. (b)$\omega_{0}=0.002$ at $t=10000$. (c) $\omega_{0}=0.02$ at $t=10000$. (d) $\omega_{0}=0.2$ at $t=10000$.}\label{1}
\end{center}
\end{figure}
\indent Figure 2 shows the snapshot of mixed chiral particles for different values of $\omega_{0}$.  Fig. 2(a) shows the distribution of particles at time $t=0$, where particles are uniformly distributed in the box and particles with oppositive chirality are mixed completely. The distributions of particles at for different $\omega_{0}$ at $t=10000$ are shown in Figs. 2(b-d).
When $\omega_{0}=0.002$, particles are still mixed (shown in Fig. 2(b)). In this case ($\omega_{0}<\Omega$), the distance between particles and obstacles is very large so that the interaction between particles and obstacles is very small. The trajectory of particles mainly depends on the speed $v_{0}$ and the rotational diffusion coefficient $D_{\theta}$.
When $\omega_{0}=0.2$ ($\omega_{0}>\Omega$), these two kinds of particles are still mixed, but repelled outside the circles whose diameter is the length $l$ of the obstacles (shown in Fig. 2(d)). In this case, the distance between particles and obstacles is very small so that interactions between particles and obstacles become very strong. For $\omega_{0}\gg\Omega$, the rotary obstacles seem like disks, particles could not enter the region of obstacles.
When $\omega_{0}=0.02=\Omega$, almost all active particles are stably captured and separated by the rotary obstacles: the CCW particles are stably distributed in the left region of the box, while the CW particles are stably stayed in the right region (shown in Fig. 2(c)). Therefore, particles with different chirality can be separated.

\begin{figure}[htbp]
\begin{center}
\includegraphics[width=8cm]{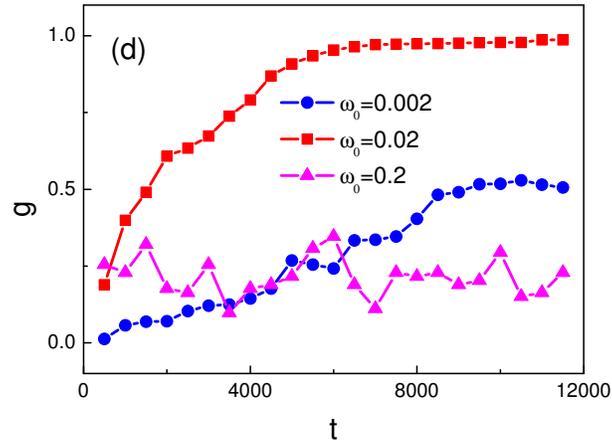}
\caption{(Color online) Gini coefficient $g$ as a function of time $t$ for different obstacle rotation values of $\omega_{0}$ at $v_{0}=0.2$ and $\Omega=0.02$. }\label{1}
\end{center}
\end{figure}

\indent To quantify particle separation, in Fig. 3 we plot the Gini coefficient $g$ as a function of time $t$ for different obstacle rotation values of $\omega_{0}$ at $v_{0}=0.2$ and $\Omega=0.02$ (the corresponding snapshot shown in Figs. 2(b-d)). It is found that the Gini coefficient $g$ is less than 1 when $\omega_0\neq \Omega$, which shows that the mixed chiral particles can not be separated even for a long time. However, when $\omega_0=\Omega$, the Gini coefficient $g$ tends to 1 after a long time, which shows that the mixed chiral particles can be completely separated. Thus we can realize the separation of particles with different chirality under the condition of $\omega_0=\Omega$.

\begin{figure}[htbp]
\begin{center}\includegraphics[width=8cm]{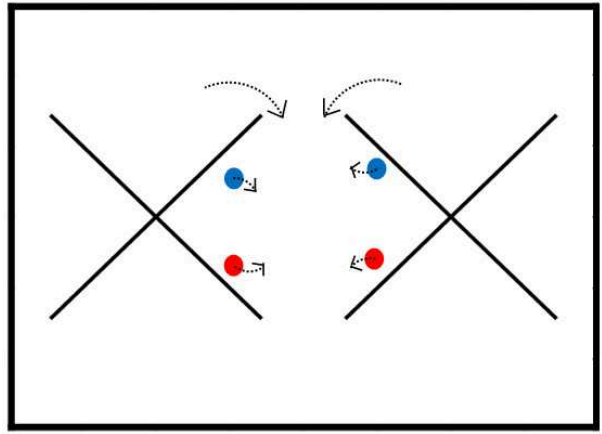}
\caption{(Color online) A schematic diagram of particles separation mechanism. The red (blue) ball represents the CCW (CW) particle.}\label{1}
\end{center}
\end{figure}

\indent In order to explain the separation mechanism, a schematic diagram is shown in Fig. 4. When the CCW particles meet the CCW rotary obstacles, they will rotate together at$\omega_0=\Omega$, therefore, the CCW particles prefer to stay in the left region of the box. However, when the CW particles meet the CCW rotary obstacles, the CW particles would be thrown out from the left region. In the same way, the CW particles will stay in the right region and the CCW particles would be thrown out from the right region. Therefore, the CCW (CW) particles will gather in the left (right) region of the box and the mixed particles can be separated.

\begin{figure}[htbp]
\begin{center}\includegraphics[width=8cm]{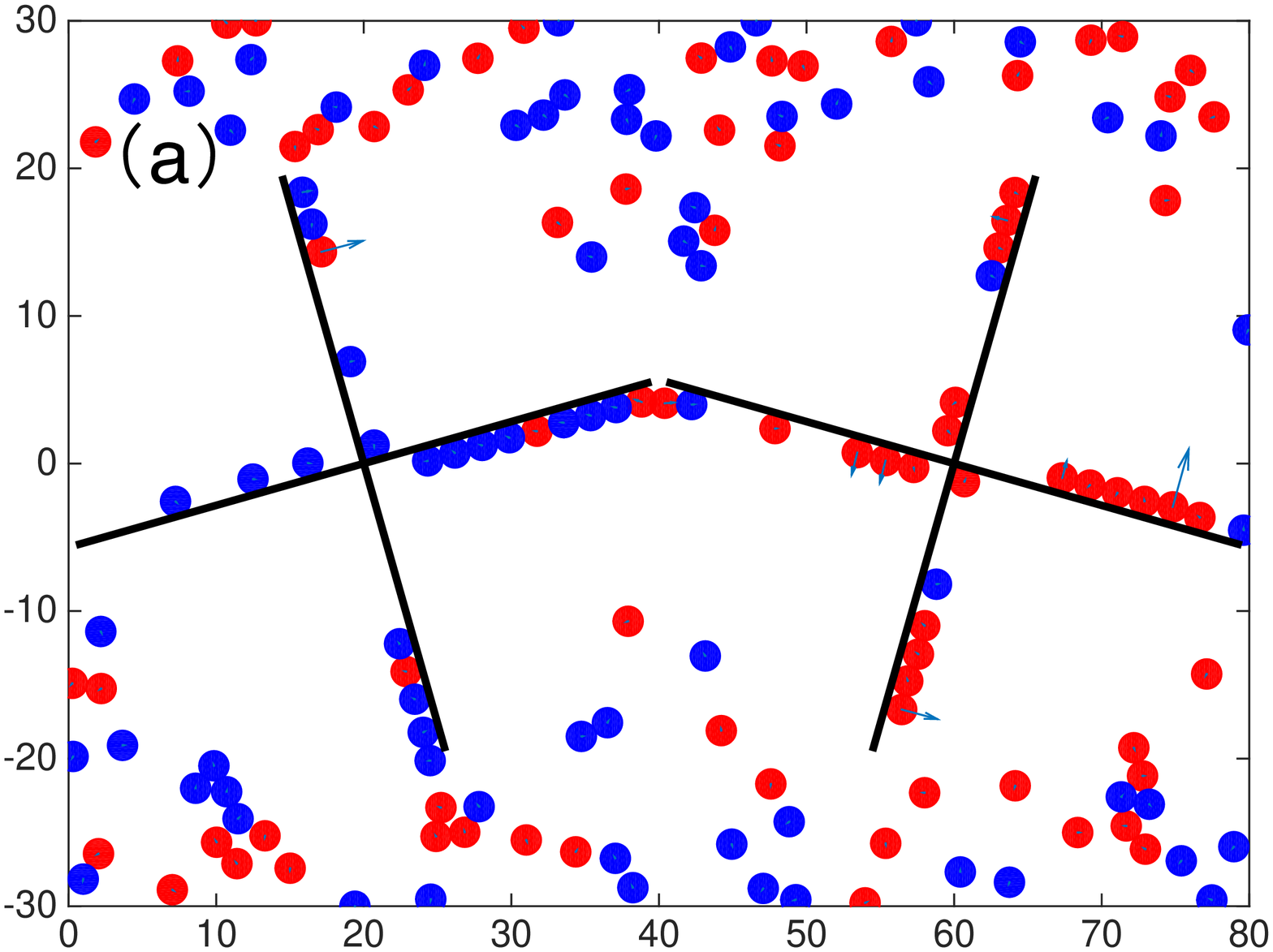}
\includegraphics[width=8cm]{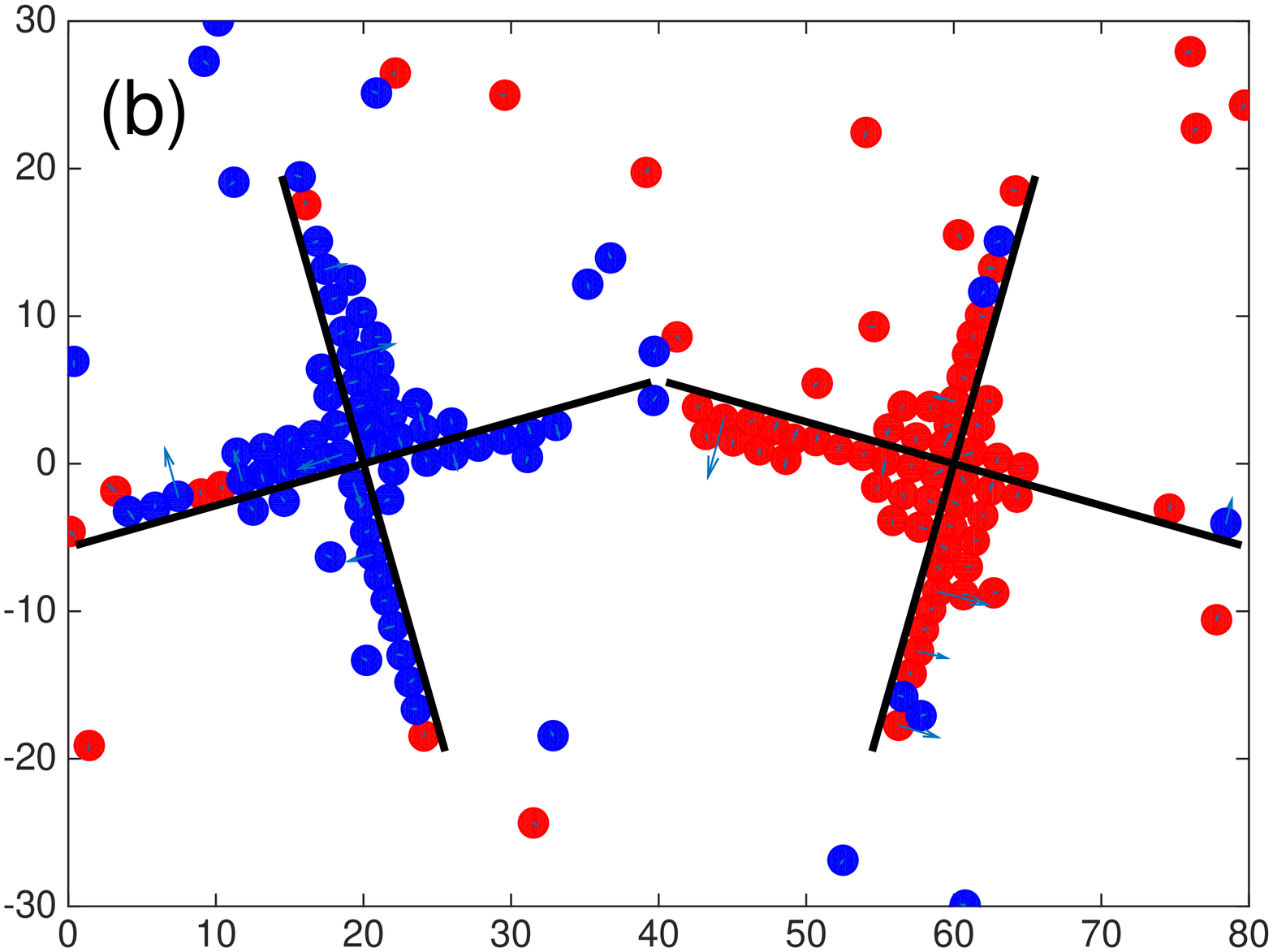}
\includegraphics[width=8cm]{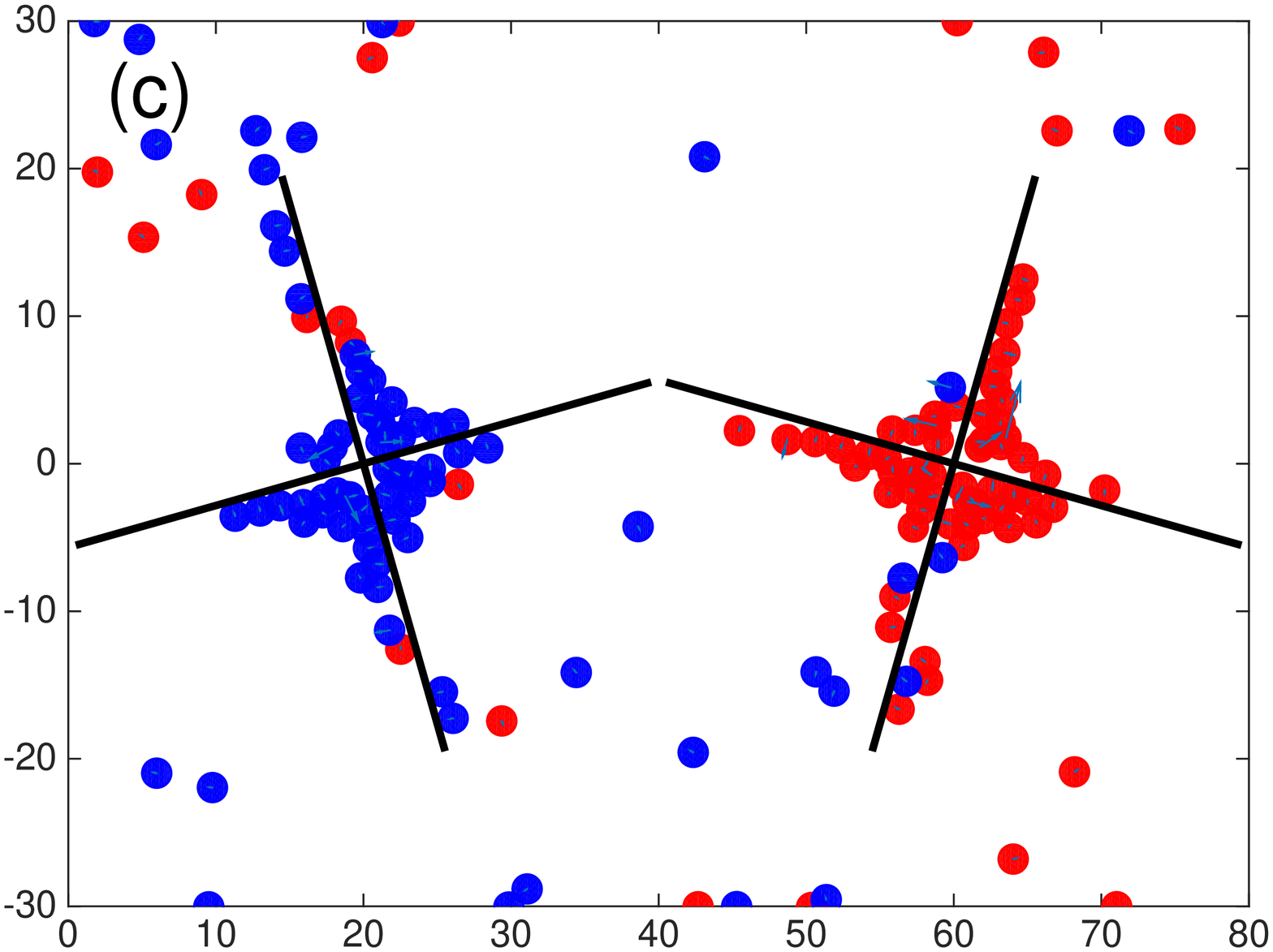}
\includegraphics[width=8cm]{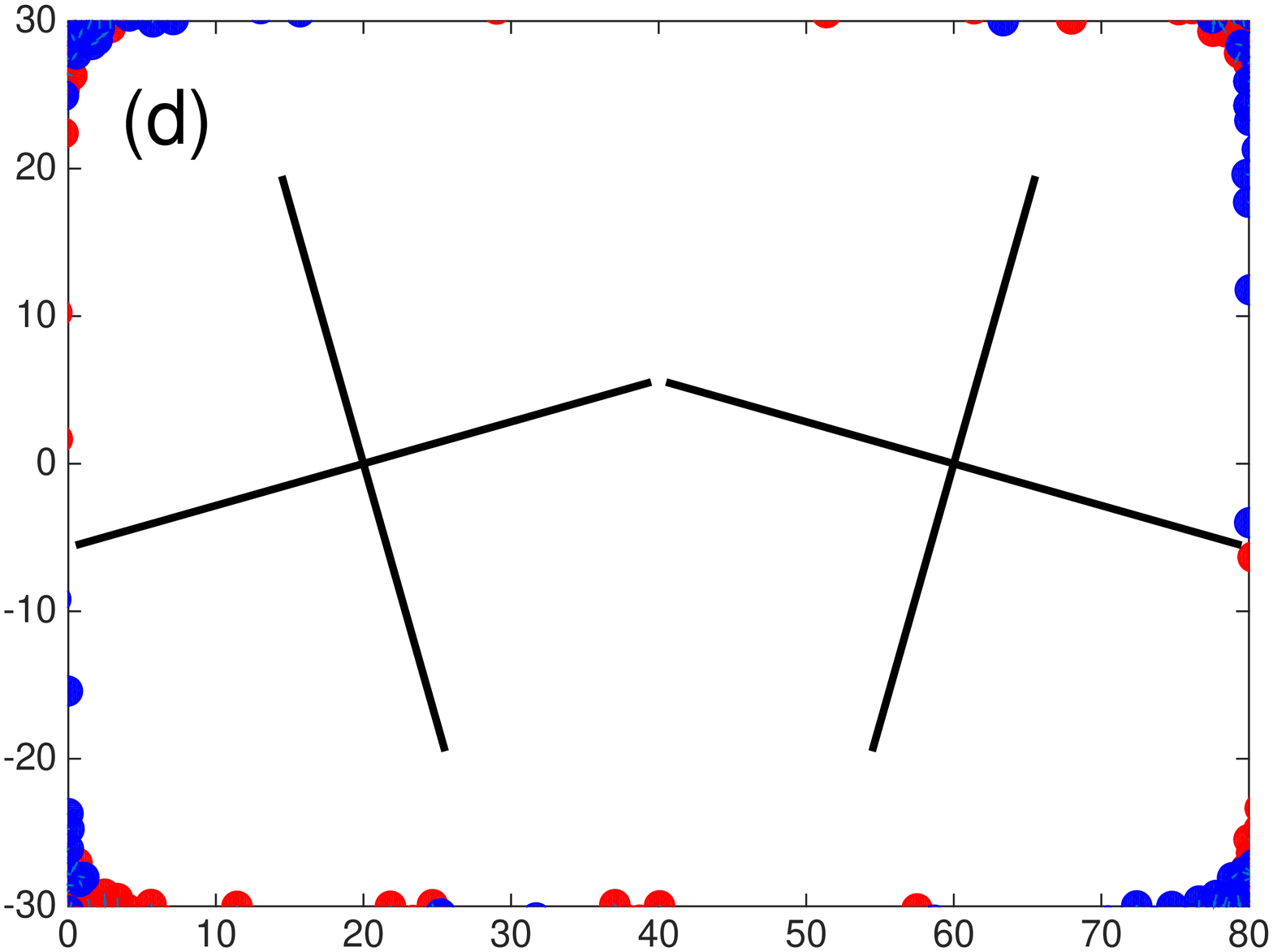}
\caption{(Color online) Snapshot of the mixed chiral particles for different values of $v_{0}$ at $\omega_{0}=\Omega=0.02$ and $t=10000$. (a) $v_{0}=0.05$. (b) $v_{0}=0.2$. (c) $v_{0}=1.0$. (d) $v_{0}=10.0$.}\label{1}
\end{center}
\end{figure}

\indent In Fig. 5, we show the effect of self-propelling speed $v_{0}$ on separating efficiency at $\omega_{0}=\Omega=0.02$. When $v_{0}=0.05$, only a few particles are captured by the obstacles, most of the particles outside the regions of obstacles are still mixed (shown in Fig. 5(a)). When $v_{0}=0.2$, almost
all particles with different chirality are gathered in different regions of rotary obstacles (shown in Fig. 5(b)). When $v_{0}=1.0$, most of the particles with different chirality are gathered in different regions of rotary obstacles (shown in Fig. 5(c)). When $v_{0}=10.0$, all particles accumulate at the boundaries of the box and no particles are trapped by the obstacles (shown in Fig. 5(d)).
\begin{figure}[htbp]
\begin{center}\includegraphics[width=8cm]{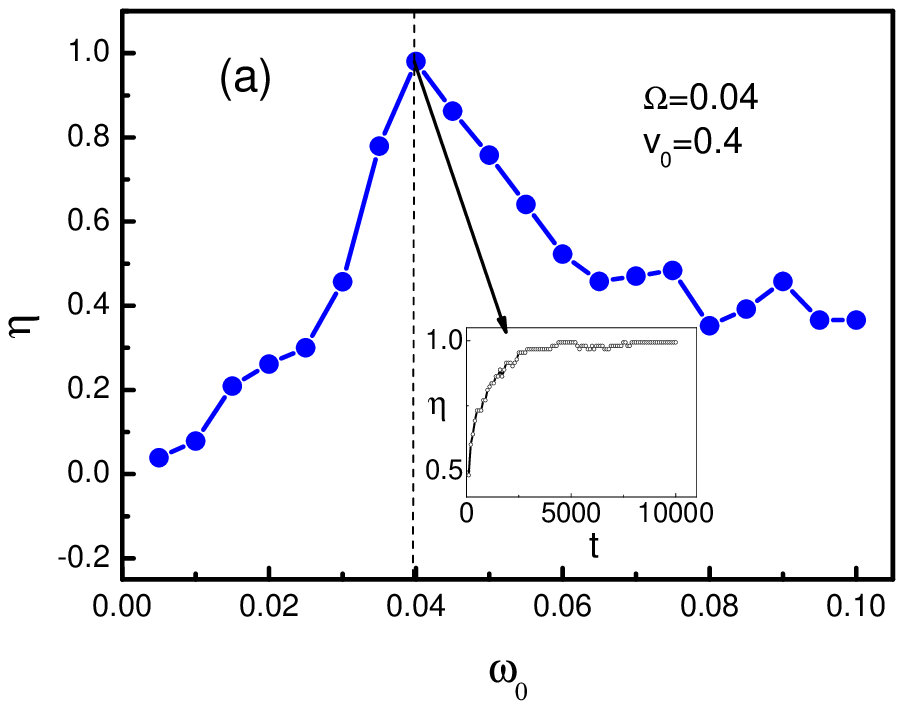}
\includegraphics[width=8cm]{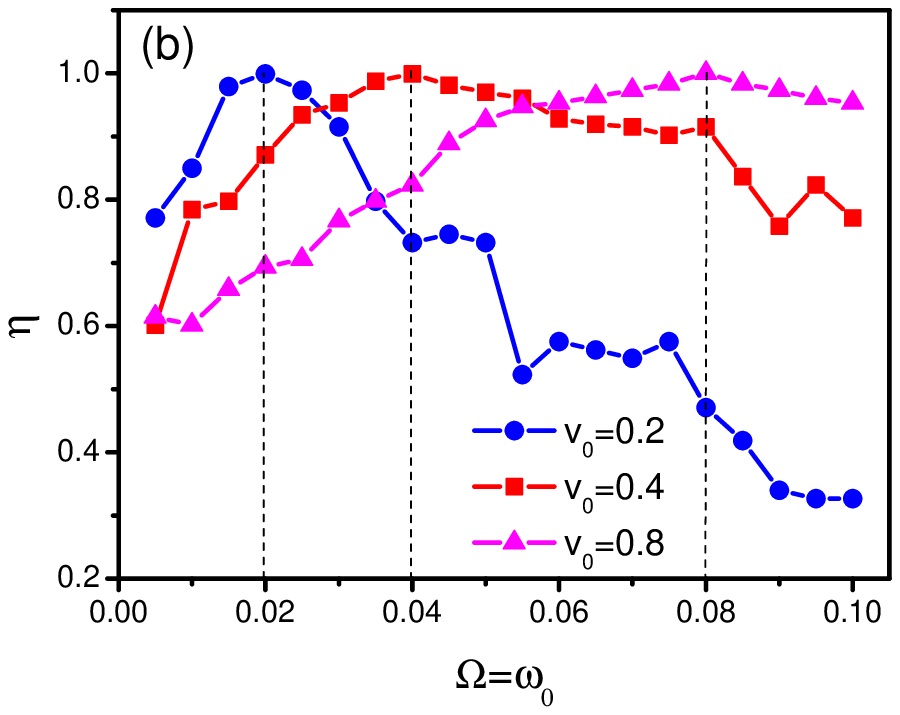}
\includegraphics[width=8cm]{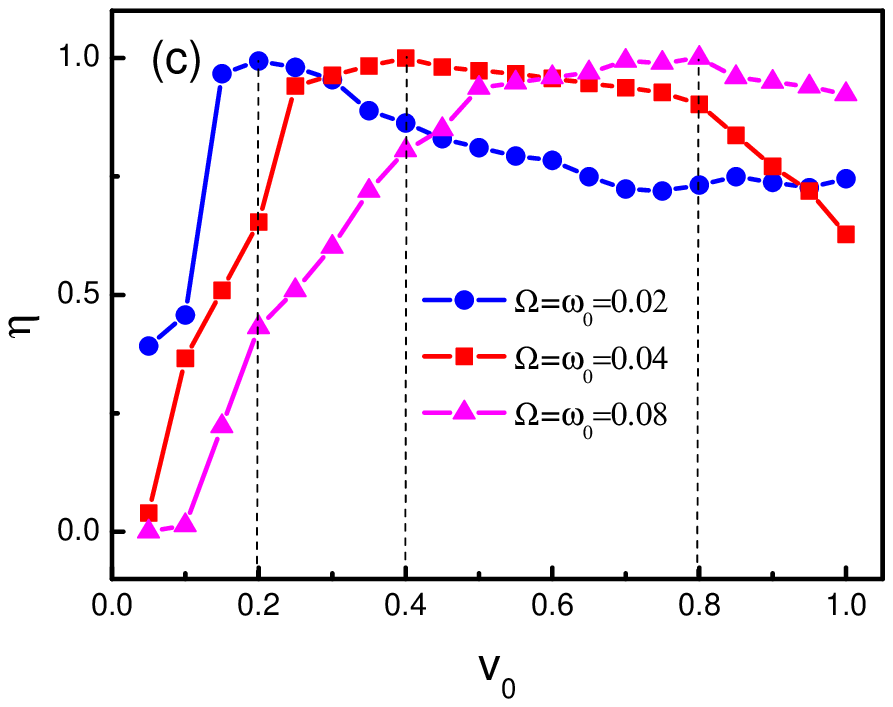}
\includegraphics[width=8cm]{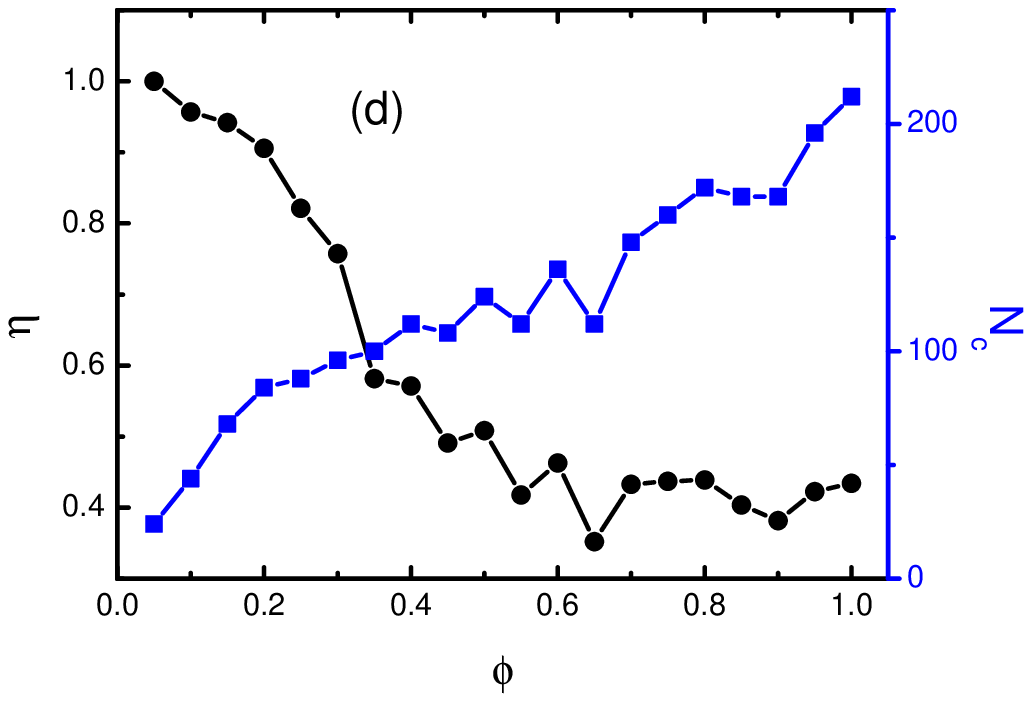}
\caption{(Color online)Capture efficiency $\eta$ of active particles at $t=50000$. (a)$\eta$ vs $\omega_0$ at $\Omega=0.04$ and $v_0=0.4$. Inset: $\eta$ vs $t$. (b)$\eta$ vs $\Omega$ at $\omega_0=\Omega$ for different values of $v_0$. (c)$\eta$ vs $v_0$ for different values of $\Omega$ and $\omega_0$. (d)$\eta$ ($N_c$) vs $\phi$ at $\Omega=\omega_0=0.04$ and $v_0=0.4$.}\label{1}
\end{center}
\end{figure}

\indent We will discuss how the parameters ($\Omega$, $\omega_0$, $v_0$ and $\phi$) affect the capture behaviors. The dependencies of the capture efficiency $\eta$ on the parameters ($\Omega$, $\omega_0$, $v_0$ and $\phi$) is shown in Fig. 6.  From Figs. 6(a) and 6(b) , we find that the capture efficiency $\eta$ tends to 1 at $\omega_0=\Omega$ after a long time (also see the inset of Fig. 6(a)). The capture efficiency $\eta$ will take its maximal value at $v_0=l\Omega/4$ (shown in Fig. 6(c)). Therefore, when the relation $\Omega=\omega_0=\frac{4v_0}{l}$ is satisfied, the mixed particles can be separated and
the capture efficiency $\eta$ goes to 1.  Fig. 6(d) shows the capture efficiency $\eta$ (the captured CW particles $N_c$ ) as a function of the packing fraction $\phi$ at $\Omega=\omega_0=\frac{4v_0}{l}$. It is found that the capture efficiency $\eta$ decreases with increasing $\phi$. This is because for large $\phi$, many CW particles can not be captured by the left rotary obstacles. However, because the total number of CW particles increase with $\phi$, $N_c$ increases with $\phi$.

\section{Concluding remarks}
\indent In this paper, we considered chiral active particles moving in the box, where there exist two rotary obstacles with different rotation directions. From numerical simulations, we mainly studied how the rotary obstacles affect the motion of chiral active particles. We found that the matching relation between the rotation angular speed $\omega_{0}$ of rotary obstacles and the angular speed $\Omega$ of the particles strongly affects separation behaviors.  When $\omega_{0}\neq\Omega$, the mixed particles were still mixed and were thrown out of the region of the obstacles. Only when $\omega_{0}=\Omega$, particles with different chirality would be captured and separated in different regions: the CCW particles would gather in the left region of the box, while the CW particles would stay in the right region. The separation efficiency also depends on the self-propelling speed $v_{0}$. When $v_{0}$ is very small, particles are mixed and could not be separated. When $v_{0}$ is very large, particles accumulate the boundaries of the box and could not be separated. There exists an optimal $v_{0}$ at which particles could be completely separated. Interestingly, we find that when the relation $\Omega=\omega_0=\frac{4v_0}{l}$ is satisfied, the mixed particles can be completely separated and the capture efficiency takes its maximal value. Our results may have application in separation and capture active particles with different chirality. Moreover, we can measure the chirality of active particles with a rotary obstacles by adapting the rotation angular speed of obstacles.

\indent  This work was supported in part by the National Natural Science Foundation of China (Grant Nos. 11575064 and 11175067), the PCSIRT (Grant No. IRT1243), the Natural Science Foundation of Guangdong Province(Grant No. 2014A030313426), the program for Excellent Talents at the University of Guangdong Province.


\begin{thebibliography}{}
\bibitem{rmp} P. H\"{a}nggi and F. Marchesoni, Rev. Mod. Phys. 81, 387(2009).
 \bibitem{Reimann}P. Reimann, Phys. Rep. 361, 57(2002).
\bibitem{Ebbens}S. J. Ebbens and J. R. Howse, Soft Matter 6, 726(2010).
\bibitem{rmp1}M. C. Marchetti, J. F. Joanny, S. Ramaswamy, T. B. Liverpool, J. Prost, M. Rao, and R. A. Simha, Rev. Mod. Phys. 85, 1143(2013).
\bibitem{rpp}J. Elgeti, R. G. Winkler, and G. Gompper, Rep. Prog. Phys. 78, 056601(2015).

\bibitem{Taylor}G. I. Taylor, Proc. Roy. Soc. London A 209, 447(1951).
\bibitem{Schweitzer}F. Schweitzer, W. Ebeling, and B. Tilch, Phys. Rev. Lett. 80, 5044(1998).
\bibitem{Erdmann}U. Erdmann, W. Ebeling, L. Schimansky-Geier, and F. Schweitzer, Eur. Phys. J. B: Condens. Matt. Comp. Sys. 15, 105(2000).
\bibitem{Toner}J.  Toner,  Y.  Tu,  and  S.  Ramaswamy,  Ann. Phys.  318, 170(2005).
\bibitem{Vicsek1}T. Vicsek and A. Zafeiris, Phys. Rep. 517, 71(2012).

\bibitem{Berg}H. C. Berg, Random Walks in Biology (Princeton University Press, Princeton, NJ, 1983).
\bibitem{Purcell} E. M. Purcell, Am. J. Phys. 45, 3(1997).
\bibitem{Howse}J. R. Howse, R. A. L. Jones, A. J. Ryan, T. Gough, R. Vafabakhsh, and R. Golestanian, Phys. Rev. Lett. 99, 048102(2007).
\bibitem{Teeffelen}S. van Teeffelen and H. L\"{o}wen, Phys. Rev. E 78, 020101(2008).
\bibitem{Leonardo}R. Di Leonardo, L. Angelani, D. Dell'Arciprete, G. Ruocco, V. Iebba, S. Schippa, M. P. Conte, F. Mecarini, F. De Angelis, and E. Di Fabrizio, Proc. Natl. Acad. Sci. USA 107, 9541(2010).
\bibitem{Aguilar}R. Ledesma-Aguilar, H. L\"{o}wen, and J. M. Yeomans, Eur. Phys. J. E 35, 70(2012).
\bibitem{Yang}Y. Yang, V. Marceau, and G. Gompper, Phys. Rev. E 82, 031904(2010).
\bibitem{Wensink}H. H. Wensink, J. Dunkel, S. Heidenreich, K. Drescher, R. E. Goldstein, H. L\"{o}wen, and J. M. Yeomans, Proc. Natl. Acad. Sci. U.S.A. 109,14308(2012).

\bibitem{H. U.}H. U. B\"{o}deker, C. Beta, T. D. Frank, and E. Bodenschatz, EPL 90, 28005(2010).
\bibitem{Friedrich} B. M. Friedrich and F. J\"{u}licher, Proc. National Acad. Sci. 104, 13256(2007).
\bibitem{Golestanian} J. R. Howse, R. A. L. Jones, A. J. Ryan, T. Gough, R. Vafabakhsh, and R. Golestanian, Phys. Rev. Lett. 99, 048102(2007).

\bibitem{Vicsek}T. Vicsek, A. Czir\'{o}k, E. Ben-Jacob, I. Cohen, and O. Shochet, Phys. Rev. Lett. 75, 1226(1995).
\bibitem{Ramaswamy}S. Ramaswamy, R. A. Simha, and J. Toner, Europhys. Lett. 62, 196(2003).
\bibitem{Peruani}F. Peruani, A. Deutsch, and M. B\"{a}r, Phys. Rev. E 74, 030904(2006).
\bibitem{Giomi}L. Giomi, T. B. Liverpool, and M. C. Marchetti, Phys. Rev. E 81, 051908(2010).
\bibitem{Saintillan}D. Saintillan, Phys. Rev. E 81, 056307(2010).
\bibitem{Shen}T. Shen and P. G. Wolynes, Proc. Natl. Acad. Sci. USA 101, 8547(2004).



\bibitem{Burada}P. S. Burada and B. Lindner, Phys. Rev. E 85,032102(2012).
\bibitem{Tailleur}J. Tailleur and M. E. Cates, Phys. Rev. Lett. 100, 218103(2008).
\bibitem{Cates}M. E. Cates, S. M. Fielding, D. Marenduzzo, E. Orlandini, and J. M. Yeomans, Phys. Rev. Lett. 101, 068102(2008).
\bibitem{Angelani}L. Angelani, A. Costanzo, and R. Di Leonardo, EPL, 96,68002(2011).
\bibitem{Fily}Y. Fily and M. C. Marchetti, Phys. Rev. Lett. 108, 235702(2012).
\bibitem{Kaiser}A. Kaiser, H. H. Wensink, and H. L\"{o}wen, Phys. Rev. Lett.108, 268307(2012).
\bibitem{Ghosh}P. K. Ghosh, V. R. Misko, F. Marchesoni, and F. Nori, Phys. Rev. Lett. 110, 268301(2013)
\bibitem{angelani}L. Angelani, R. Di Leonardo, and G. Ruocco, Phys. Rev. Lett. 102, 048104(2009).
\bibitem{Buttinoni}I. Buttinoni, J. Bialke, F. Kummel, H. L\"{o}wen, C. Bechinger, and T. Speck, Phys. Rev. Lett. 110, 238301(2013).
\bibitem{Nguyen}N. H. P. Nguyen, D. Klotsa, M. Engel, and S. C. Glotzer, Phys. Rev. Lett. 112, 075701(2014).
\bibitem{Wan}M. B. Wan, C. J. Olson Reichhardt, Z. Nussinov, and C. Reichhardt, Phys. Rev. Lett 101, 018102(2008).
\bibitem{Tailleur1}J. Tailleur and M. E. Cates, EPL 86, 60002(2009).
\bibitem{WYang}W. Yang, V. R. Misko, K. Nelissen, M. Kong, and F. M. Peeters, Soft Matter 8, 5175(2012).
\bibitem{McCandlish}S. R. McCandlish, A. Baskarana, and M. F. Hagan, Soft Matter 8, 2527(2012).
\bibitem{Maggi}C. Maggi, A. Lepore, J. Solari, A. Rizzo, and R. Di Leonardo, Soft Matter 9, 10885(2013).
\bibitem{Costanzo}A. Costanzo, J. Elgeti, T. Auth, G. Gompper, and M. Ripoll, EPL 107, 36003(2014).
\bibitem{Y}Y. Fily, A. Baskaran, and M. F. Hagan, Soft Matter 10, 5609(2014).
\bibitem{X}X. Yang, M. L. Manning, and M. C. Marchetti, Soft Matter 10, 6477(2014).
\bibitem{Berdakin}I. Berdakin, Y. Jeyaram, V. V. Moshchalkov, L. Venken, S. Dierckx, S. J. Vanderleyden, A. V. Silhanek, C. A. Condat, and V. I. Marconi,Phys. Rev. E 87, 052702(2013).
\bibitem{Reichhardt}C. Reichhardt and C. J. Olson. Reichhardt, Phys. Rev. E 88, 042306(2013).
\bibitem{Mijalkov}M. Mijalkov and G. Volpe, Soft Matter 9, 6376(2013).
\bibitem{Volpe}G. Volpe, S. Gigan, and G. Volpe, Am. J. Phys. 82, 659(2014).
\bibitem{Ai1}B. Q. Ai,Y. F. He, and W. R. Zhong, Soft Matter 11, 3852(2015).
\end{thebibliography}
\end{document}